\documentclass[letterpaper]{article} 
\usepackage[preprint]{aaai2027}  
\usepackage[hyphens]{url}  
\usepackage{graphicx} 
\usepackage{natbib}  
\usepackage{caption} 
\usepackage{algorithm}
\usepackage{algorithmic}

\usepackage{newfloat}
\usepackage{listings}
\DeclareCaptionStyle{ruled}{labelfont=normalfont,labelsep=colon,strut=off} 
\floatstyle{ruled}
\newfloat{listing}{tb}{lst}{}
\floatname{listing}{Listing}

\usepackage{booktabs}

\usepackage{makecell}
\usepackage{multirow}
\usepackage{amsmath}

\title{GroupRAG: Cognitively Inspired Group-Aware Retrieval and Reasoning via Knowledge-Driven Problem Structuring}
\author{
    Xinyi Duan\textsuperscript{\rm 1},
    Yuanrong Tang\textsuperscript{\rm 1},
    Jiangtao Gong\textsuperscript{\rm 1}\corresponding
}
\affiliations{
    \textsuperscript{\rm 1}Tsinghua University\\
    duanxy23@mails.tsinghua.edu.cn, tangxtong2022@gmail.com, gongjiangtao2@gmail.com
}

\begin{document}

\maketitle

\begin{abstract}
The performance of language models is commonly limited by insufficient knowledge and constrained reasoning. Prior approaches such as Retrieval-Augmented Generation (RAG) and Chain-of-Thought (CoT) address these issues by incorporating external knowledge or enforcing linear reasoning chains, but often degrade in real-world settings. Inspired by cognitive science, which characterizes human problem solving as search over structured problem spaces rather than single inference chains, we argue that inadequate awareness of problem structure is a key overlooked limitation. We propose GroupRAG, a cognitively inspired, group-aware retrieval and reasoning framework based on knowledge-driven keypoint grouping. GroupRAG identifies latent structural groups within a problem and performs retrieval and reasoning from multiple conceptual starting points, enabling fine-grained interaction between the two processes.  Experiments on MedQA (medical) and Bar Exam QA (legal) show that GroupRAG outperforms representative RAG- and CoT-based baselines. These results suggest that explicitly modeling problem structure, as inspired by human cognition, is a promising direction for robust retrieval-augmented reasoning.
\end{abstract}


\section{Introduction}


Language models have achieved remarkable progress across a wide range of tasks, yet they continue to struggle with complex, knowledge-dense questions that involve long contexts, multiple information sources, and intricate reasoning requirements. In medical decision making and legal case analysis, for example, relevant information is often scattered, heterogeneous, and embedded in lengthy, partially noisy descriptions.


Prior studies suggest that failures on such problems can largely be attributed to two factors: insufficient access to relevant knowledge and limited ability to reason over that knowledge. Two major lines of research have emerged to address these issues. Retrieval-Augmented Generation (RAG) incorporates external information to reduce reliance on parametric memory for knowledge-dense tasks \cite{lewis2020retrieval}. In parallel, Chain-of-Thought (CoT) prompting and related distillation methods improve reasoning by explicitly modeling intermediate inference steps \cite{wei2022chain,hsieh2023distilling}.

\begin{figure}[t]
    \centering
    \includegraphics[width=\columnwidth]{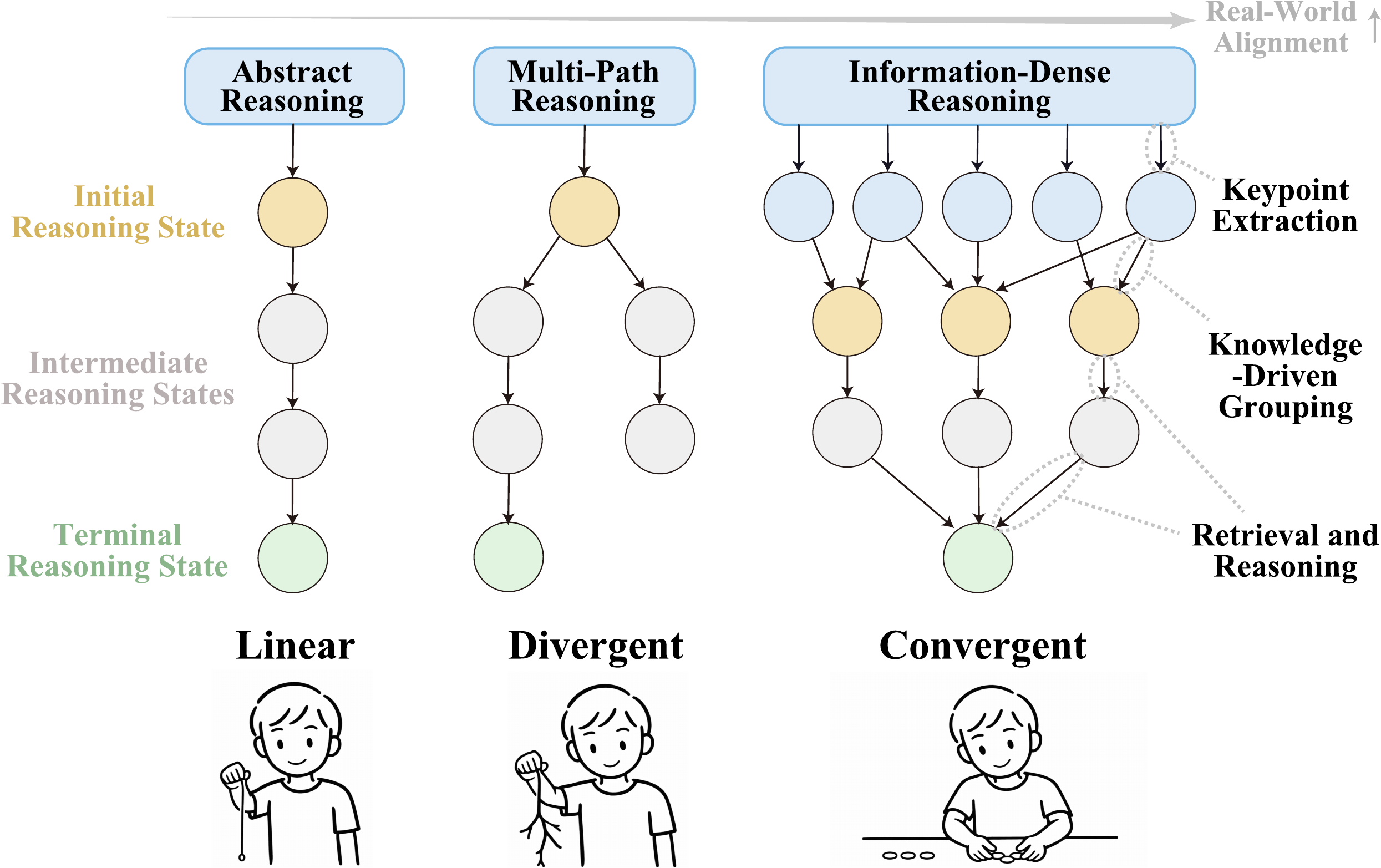}
    \caption{Reasoning Paradigms Comparison. Traditional CoT follows Linear/Divergent paths on unstructured sequences. GroupRAG transforms monolithic inputs into a structured problem space, employing a Convergent net via keypoint grouping for real-world alignment.}
    \label{figure:Head}
\end{figure}

Despite their effectiveness, existing RAG- and CoT-based approaches exhibit notable limitations in complex, real-world settings. In RAG systems, retrieved chunks often fail to precisely match the information required to answer the question, and models may struggle to align, filter, and integrate retrieved content into coherent reasoning \cite{izacard2023atlas}. CoT-based methods, while improving reasoning fluency, remain dependent on the model’s internal knowledge: when critical facts are missing or misaligned, the resulting reasoning chains may appear coherent yet rest on incomplete or incorrect premises. Consequently, simply retrieving more or reasoning longer is often insufficient for reliably solving complex problems.


Recent work has attempted to bridge this gap through structured retrieval or tighter retrieval-reasoning integration. Representative efforts include organizing external knowledge using graphs \cite{fan2025minirag,guo2024lightrag}, and interleaving retrieval steps within reasoning traces \cite{trivedi2023interleaving,wang2024rat}. While promising, these approaches often increase system complexity or still treat each question as a single undifferentiated unit for retrieval and reasoning.


A key insight of this work is that the difficulty of complex real-world questions—unlike formal mathematical or symbolic reasoning problems—lies not only in missing knowledge or insufficient reasoning capacity, but in \textit{how the problem is internally structured and represented} \cite{newell1972human,barto2013behavioral,eckstein2021mind}. Cognitive science has long shown that human problem solving depends critically on problem representation: complex tasks are understood and solved by organizing information into structured problem spaces rather than treating them as undifferentiated sequences \cite{cushen2012cues,ho2022people}. Real-world inputs are rarely monolithic. For instance, when a patient describes their condition to a physician, the narrative interleaves symptoms, medical history, test results, and irrelevant details. Yet language models process such inputs as a flat sequence, causing retrieval to operate at an inappropriate granularity. This representational mismatch in turn yields entangled and error-prone reasoning \cite{zhang2023reasoning,eckstein2021mind}.


This perspective motivates a different design principle: rather than pursuing more retrieval or longer reasoning chains, we enable reasoning over a structured problem space by uncovering the latent structure of a question. From this view, effective real-world reasoning resembles human problem solving: it begins by identifying meaningful substructures, proceeds through parallel inference from multiple conceptual starting points, and gradually converges into a coherent conclusion, as illustrated in Figure~\ref{figure:Head}. Accordingly, we focus on identifying key information points within a question and organizing them into knowledge-driven groups, each anchored to a shared knowledge concept or category, thereby providing an explicit structural scaffold for both retrieval and reasoning.


We propose \textbf{GroupRAG}, a cognitively inspired, group-aware retrieval-and-reasoning framework based on knowledge-driven keypoint grouping. GroupRAG treats grouping as a first-class operation that makes the internal structure of a question explicit, transforming unstructured inputs into structured reasoning units. Retrieval and reasoning are then performed at the group level and subsequently integrated, allowing the two processes to be temporarily decoupled yet mutually reinforcing: retrieval provides group-specific knowledge at an appropriate granularity, while reasoning over each group guides the selection and integration of relevant information toward a final answer. Our main contributions are summarized as follows:


\begin{itemize}
\item We introduce a cognitively inspired, group-aware retrieval-and-reasoning framework that explicitly models the internal structure of complex questions by organizing key information points into knowledge-driven groups, enabling retrieval and reasoning to operate at an appropriate granularity.

\item We reformulate conventional Chain-of-Thought reasoning from a single linear chain or divergent tree into a convergent reasoning net, where inference is initiated from multiple grouped reasoning roots, augmented with group-specific retrieval, and progressively integrated into a coherent global conclusion.

\item We demonstrate that GroupRAG outperforms multiple RAG-based and CoT-based methods on knowledge-intensive medical and legal question answering, showing that explicit problem structuring is critical for robust real-world reasoning.

\end{itemize}

\begin{figure*}[t]
    \centering
    \includegraphics[width=0.9\textwidth]{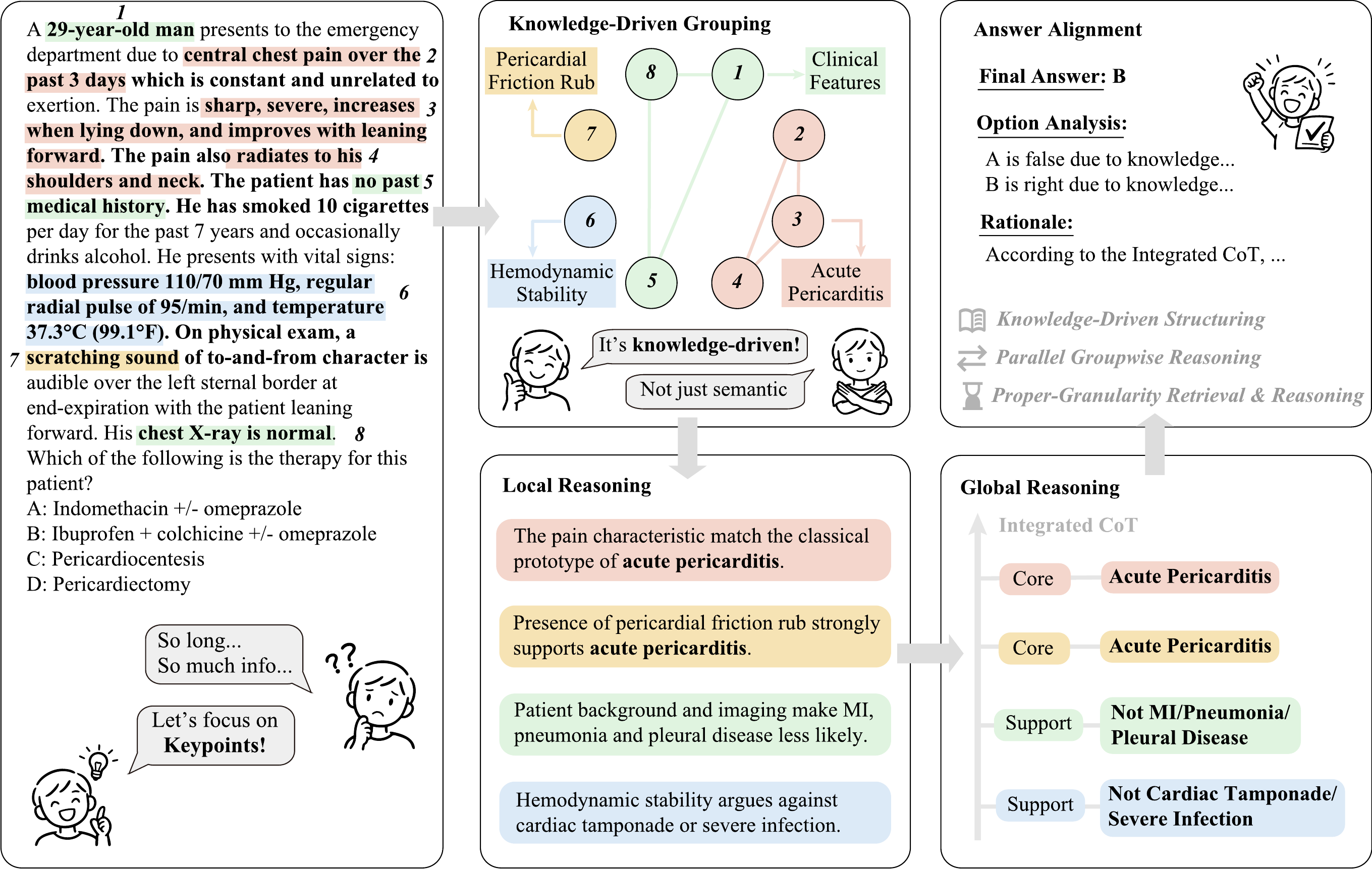}
    \caption{Illustration of the GroupRAG reasoning process on a clinical case, featuring keypoint extraction, knowledge-driven grouping, local and global reasoning, and answer alignment.}
    \label{fig:illu}
\end{figure*}

\section{Related Work}

\subsection{Chain-of-Thought and Retrieval-Augmented Generation}

Chain-of-Thought (CoT) prompting elicits multi-step reasoning in large language models by generating intermediate steps before a final answer~\cite{wei2022chain}. Subsequent work improves precision and robustness via sampling multiple reasoning paths, least-to-most decomposition, or executable code representations~\cite{wang2022selfconsistency,zhou2022least,chen2022program}. Beyond linear chains, tree, forest, and graph-structured methods explore multiple solution trajectories~\cite{yao2023tree,chen2025towards,bi2024forest,pandey2025adaptive}. However, most CoT variants still reason from a single starting point, limiting the diversity of explored paths.

Retrieval-Augmented Generation (RAG) reduces reliance on parametric memory by retrieving external knowledge for knowledge-intensive tasks~\cite{lewis2020retrieval}. Early work retrieves unstructured passages to condition generation~\cite{izacard2023atlas}, while recent methods improve retrieval quality via task-aware queries or knowledge-graph-based indices~\cite{lee2024disentangling,guo2024lightrag,fan2025minirag}. Bridging retrieval with reasoning, a growing line of work interleaves RAG with CoT, grounding intermediate steps in retrieved evidence for multi-step and domain-specific tasks~\cite{trivedi2023interleaving,wang2024rat,mavi2023retrieval,li2024rt,ma2023chain}.

\subsection{Reasoning in Cognitive Science}

Human problem solving has long been characterized in cognitive science as a process of constructing and navigating structured problem spaces, rather than following a single linear chain of inference. Classic work by Newell and Simon conceptualizes reasoning as search over an internal problem space defined by states, operators, and goals, with success depending on how the problem is represented and explored rather than on a fixed inference trajectory \cite{newell1972human}. This perspective highlights that human reasoning naturally involves multiple paths and intermediate states shaped by understanding the problem's internal structure.

Subsequent research in cognitive psychology shows that reasoning is highly sensitive to problem representation: inappropriate initial representations can lead to impasses, while restructuring or dynamically adapting task representations enables new solution paths and guides exploration \cite{barto2013behavioral,cushen2012cues,ho2022people}. These findings collectively indicate that real-world reasoning is not purely chain-based, but involves recognizing latent problem structure and exploring solutions from multiple conceptual starting points, gradually converging towards a solution.

\section{Method}

\subsection{Group-Aware Retrieval and Reasoning}

To illustrate the workflow of GroupRAG, we take complex real-world medical and legal problems as representative examples. Our goal is to model such problems in order to identify multiple starting points for reasoning chains. Inspired by how human students approach complex problems, we first employ a language model to extract key information points from the problem. This step, termed \textbf{Keypoint Extraction}, is analogous to how students highlight or circle important information in a problem.

The extracted keypoints are then organized to achieve a structured representation of the problem. We implement a \textbf{Knowledge-Driven Grouping} strategy, where a fine-tuned model leverages retrieved external knowledge to group related keypoints. This process resembles how students loop up reference materials to link strongly associated information points. After grouping, each group corresponds to a specific knowledge concept or category label. Through Keypoint Extraction and Knowledge-Driven Grouping, we achieve information structuring, transforming complex and lengthy problems into keypoint groups.

Each group is then treated as an independent starting point for reasoning. We perform groupwise retrieval and reasoning, constrained by problem conditions. This approach narrows retrieval keywords, shifting from coarse-grained (problem-level) to fine-grained (group-level) retrieval, while also limiting the reasoning scope to reduce interference from unrelated domains. The outcome is multiple \textbf{Local Reasoning} conclusions, each corresponding to a keypoint group.


These local reasoning conclusions can be categorized into three types according to their relevance and contribution to the problem: core conclusions (\textit{Core}), supporting conclusions (\textit{Support}), and noise (\textit{Noise}). Building upon this, we utilize a model to identify and select the conclusions categorized as \textit{Core} or \textit{Support}, and subsequently integrate them into a coherent global Chain-of-Thought (CoT). The CoT obtained by fusing multiple local reasoning conclusions constitutes the \textbf{Global Reasoning}.

Global Reasoning produces a readable and high-confidence reasoning chain. To align with downstream evaluation, we perform an \textbf{Answer Alignment} step. Starting from Global Reasoning, the model conducts fine-grained retrieval over candidate answer options, outputting the correct choice, option analysis and rationale. This step prevents a known failure mode in which the reasoning chain is correct but the final answer option is misaligned.

\begin{figure*}[t]
    \centering
    \includegraphics[width=0.9\textwidth]{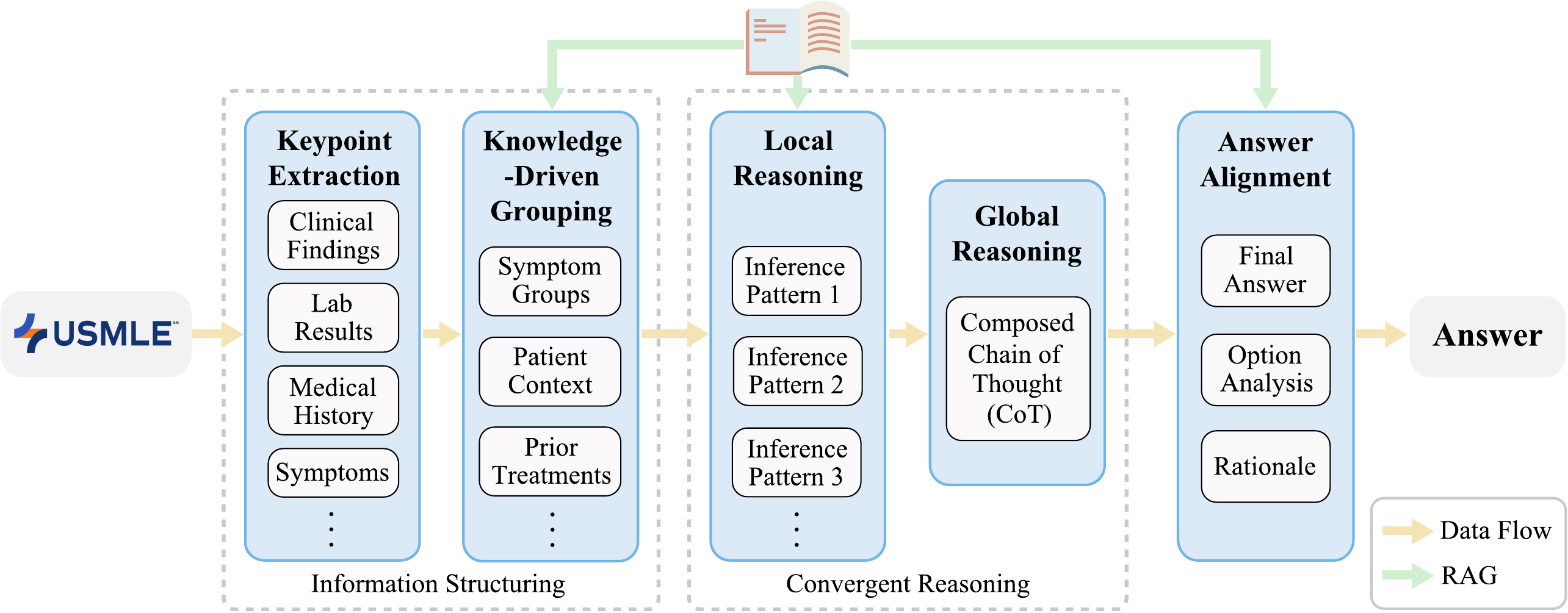}
    \caption{An Abstract Overview of GroupRAG.}
    \label{fig:method}
\end{figure*}

\subsection{System Design}

\subsubsection{Modular Pipeline and Stage-wise Training}

Building on the five-stage workflow of GroupRAG—\\
Keypoint Extraction, Knowledge-Driven Grouping, Local Reasoning, Global Reasoning, and Answer Alignment—this section presents the design details of the system. Each stage has independent inputs and outputs, functioning as sequentially connected modules.

To ensure high-quality outputs at each module, we adopt a teacher-student training paradigm. Concretely, we first run large training datasets of medical and legal questions through the complete pipeline using a large language model as the teacher, recording the inputs and outputs at each of the five modules. These records are then used as hard labels to fine-tune five dedicated student models, each specialized for one subtask. Since GroupRAG introduces a novel intermediate reasoning structure, no existing datasets provide supervision for these intermediate stages. Generating module-wise training data via a teacher model is therefore a necessary step to enable stage-wise fine-tuning.


Among these modules, the Global Reasoning module is tasked with evaluating contributions of multiple local reasoning conclusions to answering the question. This process is inherently soft and context-dependent as multiple combinations of conclusions can be valid. Supervised fine-tuning with hard labels is thus limited in capturing these nuanced dependencies. To better align the module with this task, we adopt a reinforcement learning (RL) approach, specifically utilizing a policy gradient method to fine-tune the selection policy against a custom-designed reward function.

\vspace{1ex}

\subsubsection{Global Reasoning Optimization}

The Global Reasoning module consists of two steps: a selection model identifies local reasoning conclusions that serve as \textit{Core} or \textit{Support}, and a synthesis model combines the selected conclusions into a coherent global Chain-of-Thought. Both models are independently fine-tuned, with the selection model further optimized via policy gradient to improve selection precision.

The key principle is to ensure that all \textit{Cores} are fully included, \textit{Noises} are avoided, and \textit{Supports} are encouraged. To quantitatively guide the model towards this goal, we design a reward function that captures the selection quality of local reasoning conclusions, defined as the \textit{Weighted Inference F-score (WIF)}.

\paragraph{\textit{Weighted Inference F-score (WIF).}}
Let $C$, $S$, $N$ denote the sets of Core, Support, and Noise conclusions within the local reasoning outputs. For a model-selected subset $P \subset C \cup S \cup N$, we define the recall of each category as
\[
\text{Core Recall:} \quad R_c = \frac{|P \cap C|}{|C|}\ (\text{or }1\text{ if }|C|=0)
\]
\[
\text{Support Recall:} \quad R_s = \frac{|P \cap S|}{|S|}\ (\text{or }0\text{ if }|S|=0)
\]
\[
\text{Noise Recall:} \quad R_n = \frac{|P \cap N|}{|N|}\ (\text{or }0\text{ if }|N|=0)
\]
The WIF is then computed as:
\[
\text{WIF}(P) = R_c^\alpha \cdot (1 - R_n)^\beta \cdot (1 + \gamma R_s)
\]
where $\alpha$ controls the reward weight for covering Cores, $\beta$ controls the penalty for selecting Noises, and $\gamma$ provides mild encouragement for including Supports, with $\alpha \ge \beta \gg \gamma$ reflecting the priority: Core coverage $>$ Noise avoidance $\gg$ Support inclusion. In practice, we tested a range of values and found that $\alpha = 2.5,\ \beta = 2,\ \gamma = 0.5$ yield the best global reasoning quality. If $P=\emptyset$, we assign $\text{WIF}(P)=0$ to discourage the model from skipping all conclusions.

\paragraph{\textit{Policy Optimization.}}
We model selection as a stochastic policy $\pi_\theta(P \mid x)$, parameterized by independent Bernoulli distributions over each local conclusion: for each conclusion $l_i$, the model outputs a selection probability $p_i$, and a rollout $P_k$ is generated by sampling each $l_i$ independently. For each problem, we draw $K = 5$ rollouts $\{P_k\}_{k=1}^{K}$, compute their WIF rewards $R_k = \text{WIF}(P_k)$, and estimate advantages as normalized rewards $\hat{A}_k = \frac{R_k - \bar R}{\text{std}(\{R_i\})+\epsilon}$, where $\bar R$ is the mean reward over rollouts. The policy is updated via policy gradient $\mathcal{L} = -\frac{1}{K} \sum_{k=1}^{K} \hat{A}_k \cdot \log \pi_\theta(P_k \mid x)$. This training procedure ensures that the model learns to select local reasoning conclusions that maximize WIF, producing a coherent global Chain-of-Thought that prioritizes \textit{Core} while avoiding \textit{Noise}.

\subsection{Retrieval-Augmented Generation in GroupRAG}

Retrieval-Augmented Generation (RAG) plays a crucial role in GroupRAG and is integrated into three key modules: Knowledge-Driven Grouping, Local Reasoning, and Answer Alignment. Rather than serving as a standalone retrieval component, RAG is tightly coupled with the groupwise reasoning structure and operates at different granularities across the pipeline.

In Knowledge-Driven Grouping, RAG is applied at the keypoint level. For each extracted keypoint, the model retrieves external knowledge to ground it in a relevant knowledge context. When multiple keypoints retrieve overlapping or highly related knowledge, they are inferred to share the same underlying domain concept and are grouped together accordingly, ensuring that keypoints within each group relate to a specific knowledge concept rather than being clustered by surface-level semantic similarity.

In Local Reasoning, RAG operates at the group level. Each group of keypoints is treated as a unified query for retrieval, with the objective of identifying knowledge that jointly explains multiple keypoints. For example, retrieving information for two symptoms independently may lead to different candidate diseases, whereas retrieving their combination may reveal that they are associated symptoms of the same disease. By conditioning retrieval on grouped keypoints, the model is able to perform more precise and context-aware reasoning. RAG in Answer Alignment is demonstrated before in this section. 

Overall, RAG in GroupRAG is adapted to different stages of the reasoning process, operating at varying granularities from the keypoint level to the group level and the option level. This group-aware integration of retrieval enables GroupRAG to progressively refine both the granularity and relevance of retrieved knowledge across the reasoning pipeline.

\section{Experiments}

\subsection{Dataset and Model}

To evaluate the effectiveness of GroupRAG in addressing complex real-world reasoning problems, we adopt two knowledge-intensive datasets from distinct domains: MedQA\cite{jin2021disease} for medical question answering and Bar Exam QA\cite{zheng2025barexam} for legal question answering. MedQA consists of long and information-dense USMLE-style clinical case descriptions that require multi-step reasoning over a wide range of topics in basic and clinical medicine. Bar Exam QA comprises Multistate Bar Examination (MBE) questions, each presenting a detailed legal scenario with four answer choices and demanding analogical reasoning across statutes and case precedents. Both datasets share the same core challenge: key facts are scattered across lengthy, information-dense, and partially noisy contexts and must be identified, organized, and synthesized through multi-step reasoning. Because medicine and law share no surface vocabulary or domain-specific patterns, consistent gains across both datasets verify that GroupRAG addresses this shared cognitive challenge rather than fitting to a single domain.

From each dataset, we randomly select 5,000 questions as the training set. Following the procedure described in System Design, each training question is sequentially processed by the five modules of GroupRAG, each instantiated with GPT-4o\cite{openai2024gpt4o}, with the module-wise outputs recorded as intermediate supervision signals for training. For evaluation, we construct separate test sets of 1,000 questions per dataset using stratified sampling: questions are first categorized into complex and basic pools based on physical metrics such as entity counts and text lengths, then 80\% are drawn from the complex pool as a stress test and 20\% from the basic pool as anchor cases. This distribution verifies both GroupRAG's capacity for complex reasoning and its graceful degradation on simpler queries.

We choose LLaMA3.1-8B\cite{dubey2024LLaMA} as the base model. State-of-the-art LLMs are closed-source and do not support full-parameter SFT, making open-source SLMs the practical choice for module-wise training; validating reasoning frameworks on 7B-8B models is also standard practice in this field~\cite{wang2025rare, zhang2025imprag}. Using the collected intermediate data, the base model is separately fine-tuned on each dataset, producing two independent sets of five lightweight sub-models. This cross-domain design allows us to verify that the pipeline architecture itself, rather than any domain-specific advantage, drives the observed improvements.

\subsection{Evaluation Metrics}

We design stage-wise evaluation metrics for the five modules of GroupRAG.

\paragraph{Keypoint Extraction.}
Extraction performance is evaluated using precision, recall, and F1 score. For each question, the keypoints extracted by the trained lightweight model are compared against a gold standard set of keypoints extracted by GPT-4o. Precision is defined as the proportion of predicted keypoints that correctly match the gold standard keypoints, while recall measures the proportion of gold standard keypoints that are successfully recovered. The F1 score, computed as the harmonic mean of precision and recall, serves as the primary metric because the extraction task requires simultaneously maximizing coverage of gold keypoints (recall) and minimizing spurious outputs (precision).

\paragraph{Knowledge-Driven Grouping.}
The grouping stage aims to partition the extracted keypoints into groups associated with the same pieces of knowledge, which naturally constitutes a clustering task rather than a classification task. We adopt BCubed F1 as the primary metric because it evaluates clustering quality at the level of individual keypoints rather than global partition matching (e.g., NMI or ARI), which directly aligns with the grouping objective. BCubed Precision measures, for each keypoint, the proportion of other keypoints in the same predicted group that also belong to the same gold-standard group generated by GPT-4o. BCubed Recall measures the proportion of keypoints in the gold-standard group that are correctly placed into the same predicted group. BCubed F1 is computed as the harmonic mean of BCubed Precision and BCubed Recall, and the final score is obtained by averaging over all keypoints.

\paragraph{Local and Global Reasoning.}
For local reasoning based on each keypoint group, we use GPT-4o to evaluate the factual and logical correctness of each inference, and then compute the overall accuracy over all inferences. For global reasoning, we adopt the WIF function to assess the model’s ability to distinguish Core, Support, and Noise local conclusions. This ensures that the assembled global reasoning covers all essential conclusions while filtering out distracting or irrelevant inferences.

\paragraph{Answer Alignment.}
To evaluate the final output of GroupRAG, we compare the selected answer options with the correct options in the dataset, and calculate the overall accuracy across the test set. This accuracy serves as the primary metric for horizontal comparison with other models and methods.

\subsection{Experimental Design}
To systematically evaluate the effectiveness of GroupRAG and analyze the contribution of its individual modules, we design three sets of experiments: leave-one-out ablation, progressive ablation, and joint comparison across different models and methods. All experiments are conducted over 5 independent runs on the test set, and results are reported as averages.

\paragraph{Leave-One-Out Ablation.}
In this setting, we assess the marginal contribution of each module in GroupRAG by removing one module at a time while keeping all other components unchanged. Specifically, for each of the five modules, we either replace the specially trained model with the base model, or remove the RAG component within the module, if it originally contains one. Each ablation variant is evaluated using the same set of metrics as the complete GroupRAG system. The performance of each variant is then compared against the complete GroupRAG pipeline, enabling a fine-grained analysis of the individual impact of each module.

\paragraph{Progressive Ablation.}
While leave-one-out ablation focuses on isolated effects, progressive ablation is designed to examine the cumulative contribution of GroupRAG’s modules. Starting from the complete GroupRAG system, we progressively replace trained models with the base model and sequentially remove RAG modules in the order they appear in the pipeline. This process continues until the system degenerates into a baseline configuration composed entirely of base models and without any RAG component. Each experimental setting is compared with the preceding one, allowing us to observe how performance evolves as modules are incrementally removed.

\paragraph{Joint Comparison Across Models and Methods.}
In the final set of experiments, we aim to evaluate the capability gap of small language models under different reasoning and retrieval paradigms, and to compare their performance with a reference model. To this end, we conduct a horizontal comparison across different models and methods, evaluating LLM, SLM, and trained SLM under CoT prompting, standard RAG, and GroupRAG. Specifically, for the trained SLM setting, models are separately fine-tuned under different supervision signals corresponding to each method, such as question-answer pairs, CoT, and the intermediate data of GroupRAG. For the LLM setting, GPT-4o is included as a reference model to provide an approximate upper bound on task performance. For clarity and consistency, we focus on final answer accuracy as the sole evaluation metric in this comparison.

\begin{table}
    \centering
    \footnotesize
    \setlength{\tabcolsep}{3pt}
    \resizebox{\columnwidth}{!}{
    \begin{tabular}{l *{5}{cc}}
        \toprule
          & \multicolumn{2}{c}{\makecell[c]{Extract\\F1}} 
          & \multicolumn{2}{c}{\makecell[c]{Group\\F1}} 
          & \multicolumn{2}{c}{\makecell[c]{Local\\Acc.(\%)}} 
          & \multicolumn{2}{c}{\makecell[c]{Global\\WIF}} 
          & \multicolumn{2}{c}{\makecell[c]{Answer\\Acc.(\%)}} \\
        \cmidrule(lr){2-3} \cmidrule(lr){4-5} \cmidrule(lr){6-7} \cmidrule(lr){8-9} \cmidrule(lr){10-11}
          & Med & Bar & Med & Bar & Med & Bar & Med & Bar & Med & Bar \\
        \midrule
        GroupRAG    & \textbf{0.96} & \textbf{0.92} & \textbf{0.80} & \textbf{0.74} & \textbf{73.1} & \textbf{74.7} & \textbf{1.13} & \textbf{1.02} & \textbf{71.7} & \textbf{66.4} \\
        w/o Ext. T. & \textbf{0.95} & \textbf{0.90} & 0.81 & 0.73          & 70.2          & 71.8          & 1.06          & 0.95          & 68.5          & 62.1          \\
        w/o Gro. T. & 0.96          & 0.92          & \textbf{0.71} & \textbf{0.65} & 64.5          & 67.9          & 0.93          & 0.87          & 63.0          & 59.7          \\
        w/o Loc. T. & 0.96          & 0.92          & 0.80          & 0.74          & \textbf{61.8} & \textbf{63.1} & 0.85          & 0.76          & 64.2          & 62.5          \\
        w/o Glo. T. & 0.96          & 0.92          & 0.80          & 0.74          & 73.1          & 74.7          & \textbf{0.69} & \textbf{0.67} & 68.2          & 63.7          \\
        w/o Ans. T. & 0.96          & 0.92          & 0.80          & 0.74          & 73.1          & 74.7          & 1.13          & 1.02          & \textbf{67.5} & \textbf{62.9} \\
        \midrule
        w/o Gro. R. & 0.96          & 0.92          & \textbf{0.70} & \textbf{0.62} & 62.8          & 66.7          & 0.91          & 0.84          & 64.5          & 60.1          \\
        w/o Loc. R. & 0.96          & 0.92          & 0.80          & 0.74          & \textbf{59.5} & \textbf{61.7} & 0.86          & 0.77          & 63.2          & 58.1          \\
        w/o Ans. R. & 0.96          & 0.92          & 0.80          & 0.74          & 73.1          & 74.7          & 1.13          & 1.02          & \textbf{67.2} & \textbf{61.9} \\
        \bottomrule
    \end{tabular}
    }
    \caption{Leave-One-Out Ablation Results. Each variant is compared with the full pipeline (first row) to demonstrate the marginal contribution of each module. ``w/o'' indicates removing each component independently from the full pipeline (T.: Training; R.: RAG; Med: MedQA; Bar: Bar Exam QA).}
    \label{tab:leave}
\end{table}

\section{Results}

\paragraph{Leave-One-Out Ablation.}

Results are shown in Table~\ref{tab:leave}. Since the five modules are executed sequentially, ablating a specific module only affects its own evaluation metric and those of downstream modules, while leaving upstream metrics unchanged. Compared with the complete GroupRAG system, all ablation variants exhibit varying degrees of degradation in final answer accuracy across both datasets, indicating that each of the five trained modules and their associated RAG components plays a role in the overall performance of GroupRAG. A closer comparison across ablation variants shows that four configurations experience the largest drops in answer accuracy, corresponding to the removal of trained models and RAG components in the Knowledge-Driven Grouping and Local Reasoning modules (7-9\% on MedQA, 4-8\% on Bar Exam QA). In contrast, ablating the Keypoint Extraction and Global Reasoning modules leads to relatively smaller decreases in final accuracy (approximately 3-5\%). The ranking of module importance is preserved across domains.

\paragraph{Progressive Ablation.}

Results are shown in Table~\ref{tab:progressive}. As modules are removed sequentially following the pipeline order, evaluation metrics of downstream stages exhibit a gradual decline across both datasets, forming a stage-wise degradation pattern that aligns with the pipeline structure. As fewer modules are retained, degradation in upstream outputs and intermediate reasoning quality accumulates and results in a monotonic decrease in final answer accuracy. Among all transitions, removing the trained models of the Grouping and Local Reasoning modules leads to the largest accuracy declines in both domains, corroborating the leave-one-out findings.

\begin{table}
    \centering
    \footnotesize
    \setlength{\tabcolsep}{3pt}
    \resizebox{\columnwidth}{!}{
    \begin{tabular}{l *{5}{cc}}
        \toprule
          & \multicolumn{2}{c}{\makecell[c]{Extract\\F1}} 
          & \multicolumn{2}{c}{\makecell[c]{Group\\F1}} 
          & \multicolumn{2}{c}{\makecell[c]{Local\\Acc.(\%)}} 
          & \multicolumn{2}{c}{\makecell[c]{Global\\WIF}} 
          & \multicolumn{2}{c}{\makecell[c]{Answer\\Acc.(\%)}} \\
        \cmidrule(lr){2-3} \cmidrule(lr){4-5} \cmidrule(lr){6-7} \cmidrule(lr){8-9} \cmidrule(lr){10-11}
          & Med & Bar & Med & Bar & Med & Bar & Med & Bar & Med & Bar \\
        \midrule
        GroupRAG   & \textbf{0.96} & \textbf{0.92} & \textbf{0.80} & \textbf{0.74} & \textbf{73.1} & \textbf{74.7} & \textbf{1.13} & \textbf{1.02} & \textbf{71.7} & \textbf{66.4} \\
        $-$Ext. T. & \textbf{0.95} & \textbf{0.90} & 0.81          & 0.73          & 70.2          & 71.8          & 1.06          & 0.95          & 68.5           & 63.5           \\
        $-$Gro. T. & 0.95          & 0.90          & \textbf{0.72} & \textbf{0.67} & 62.1          & 63.9          & 0.94          & 0.84          & 63.0           & 58.7           \\
        $-$Gro. R. & 0.95          & 0.90          & \textbf{0.70} & \textbf{0.63} & 61.8          & 62.1          & 0.88          & 0.81          & 61.0           & 57.3           \\
        $-$Loc. T. & 0.95          & 0.90          & 0.70          & 0.63          & \textbf{58.5} & \textbf{58.1} & 0.75          & 0.71          & 56.7           & 53.2           \\
        $-$Loc. R. & 0.95          & 0.90          & 0.70          & 0.63          & \textbf{55.7} & \textbf{54.2} & 0.69          & 0.63          & 56.0           & 52.9           \\
        $-$Glo. T. & 0.95          & 0.90          & 0.70          & 0.63          & 55.7          & 54.2          & \textbf{0.55} & \textbf{0.52} & 53.7           & 50.7           \\
        $-$Ans. T. & 0.95          & 0.90          & 0.70          & 0.63          & 55.7          & 54.2          & 0.55          & 0.52          & \textbf{52.7}  & \textbf{49.3}  \\
        $-$Ans. R. & 0.95          & 0.90          & 0.70          & 0.63          & 55.7          & 54.2          & 0.55          & 0.52          & \textbf{51.0}  & \textbf{48.8}  \\
        \bottomrule
    \end{tabular}
    }
    \caption{Progressive Ablation Results. Each variant is compared with the previous row to demonstrate the cumulative effect of the integrated modules. ``$-$'' indicates cumulative removal (T.: Training; R.: RAG; Med: MedQA; Bar: Bar Exam QA).}
    \label{tab:progressive}
\end{table}

\paragraph{Joint Comparison Across Models and Methods.}

Results are shown in Table~\ref{tab:joint_compare}. Across both datasets, for the untrained base model LLaMA3.1-8B, CoT prompting and Naive RAG yield moderate accuracy improvements over the direct baseline, whereas GroupRAG leads to substantially larger gains. Across all retrieval and reasoning methods, the trained LLaMA3.1-8B consistently outperforms its untrained counterpart. Among small-model configurations, applying GroupRAG to the trained SLM achieves the highest accuracy, reaching 71.7\% on MedQA and 66.4\% on Bar Exam QA. GPT-4o also benefits from GroupRAG on both datasets, though the relative gain over direct answering is markedly smaller than for SLMs.

\begin{table}
    \centering
    \footnotesize
    \setlength{\tabcolsep}{2.5pt}
    \resizebox{\columnwidth}{!}{
    \begin{tabular}{llrrrr}
        \toprule
        \multirow{2}{*}{Dataset}
        & \multirow{2}{*}{Model}
        & \multicolumn{4}{c}{Method} \\
        \cmidrule(lr){3-6}

        &
        & \makecell[c]{Direct\\(\%)}
        & \makecell[c]{+CoT\\Prompting(\%)}
        & \makecell[c]{+Naive\\RAG(\%)}
        & \makecell[c]{+Group\\RAG(\%)} \\
        \midrule

        \multirow{3}{*}{MedQA}
        & GPT-4o           & $\mathbf{89.1}_{\pm 0.8}$ & $90.3_{\pm 0.7}$ & $88.7_{\pm 0.7}$ & $\mathbf{91.5}_{\pm 0.5}$ \\
        & L3.1-8B(base)    & $\mathbf{48.1}_{\pm 1.0}$ & $54.4_{\pm 1.0}$ & $53.9_{\pm 1.1}$ & $\mathbf{61.0}_{\pm 0.8}$ \\
        & L3.1-8B(trained) & $\mathbf{52.9}_{\pm 0.6}$ & $61.3_{\pm 0.9}$ & $59.1_{\pm 0.8}$ & $\mathbf{71.7}_{\pm 0.7}$ \\

        \midrule

        \multirow{3}{*}{BarExam}
        & GPT-4o           & $\mathbf{77.5}_{\pm 0.7}$ & $79.3_{\pm 0.8}$ & $78.2_{\pm 0.9}$ & $\mathbf{80.7}_{\pm 0.8}$ \\
        & L3.1-8B(base)    & $\mathbf{42.9}_{\pm 1.0}$ & $48.6_{\pm 0.8}$ & $47.8_{\pm 0.9}$ & $\mathbf{55.0}_{\pm 0.9}$ \\
        & L3.1-8B(trained) & $\mathbf{48.8}_{\pm 0.6}$ & $55.5_{\pm 0.8}$ & $53.5_{\pm 0.8}$ & $\mathbf{66.4}_{\pm 0.7}$ \\

        \bottomrule
    \end{tabular}
    }
    \caption{Joint comparison across datasets, models, and methods. All values report mean accuracy over 5 independent runs with standard deviation. L3.1-8B refers to LLaMA 3.1-8B.}
    \label{tab:joint_compare}
\end{table}

\section{Discussion}

Results of the leave-one-out ablation study indicate that properly uncovering problem structure and conducting local reasoning are central to accurate problem solving. In contrast, tasks that are more procedural in nature—for example, extracting information—can be properly handled by untrained small models, and thus have limited impact on the final answer accuracy. Results of the progressive ablation study demonstrate that the performance gain of GroupRAG arises from the cumulative synergy between modules. The quality of upstream outputs directly affects subsequent reasoning, and any upstream degradation is amplified through the pipeline, ultimately impacting the final answer.

Joint comparison across models and methods indicates that GroupRAG effectively compensates for the knowledge and reasoning limitations of small language models, enabling them to answer complex questions more accurately and robustly. For GPT-4o, GroupRAG yields only modest gains over direct answering. This can be attributed to two factors: first, large language models already possess strong inherent knowledge coverage and implicit reasoning capabilities, leaving less room for external augmentation; second, and more importantly, the GroupRAG pipeline applied to GPT-4o is not fine-tuned with module-wise training, whereas the ablation studies confirm that dedicated module-level training is a critical component of GroupRAG's overall effectiveness.

Beyond quantitative results, we also examined GroupRAG's behavior on simpler queries from the test set. When a question involves only a few keypoints or limited knowledge demands, the Knowledge-Driven Grouping module naturally organizes all extracted keypoints into a single group. In such cases, GroupRAG gracefully degenerates into standard RAG with linear CoT reasoning over a single starting point, and its final accuracy remains no worse than the corresponding baseline, ensuring robustness across the full spectrum of query complexity.

Future work could proceed in two directions. First, the current instantiation relies on a handcrafted set of five modules with a fixed topology. A natural generalization is to move beyond this predetermined configuration toward a more adaptive architecture, where the set and connectivity of reasoning modules are dynamically composed based on the structure of each problem, rather than following a one-size-fits-all pipeline. Second, more sophisticated methods for modeling internal problem structures could better constrain retrieval and reasoning, thereby improving both accuracy and robustness across tasks and model scales.

\section{Conclusion}


In this paper, we proposed GroupRAG, a cognitively inspired framework that models the internal structure of complex questions by organizing key information points into knowledge-driven groups and reformulating CoT into a convergent reasoning net over multiple grouped reasoning roots. By decoupling retrieval and reasoning and re-coupling them at the group level, GroupRAG enables fine-grained interaction between the two processes. Empirical evaluations on medical and legal QA demonstrate consistent gains over representative RAG- and CoT-based methods. Ultimately, this work highlights that explicitly modeling problem structure is a promising direction for robust real-world reasoning, beyond the mere extension of reasoning chains.

\newpage

\bibliography{aaai2027}

@article{jin2021disease,
  title={What disease does this patient have? a large-scale open domain question answering dataset from medical exams},
  author={Jin, Di and Pan, Eileen and Oufattole, Nassim and Weng, Wei-Hung and Fang, Hanyi and Szolovits, Peter},
  journal={Applied Sciences},
  volume={11},
  number={14},
  pages={6421},
  year={2021},
  publisher={MDPI}
}

@misc{dubey2024llama,
  title={The Llama 3 Herd of Models},
  author={Dubey, Abhimanyu and Jauhri, Abhinav and Pandey, Abhinav and Kadian, Abhishek and Al-Dahle, Ahmad and Letman, Aiesha and Mathur, Akhil and Schelten, Alan and Yang, Amy and Fan, Angela and others},
  year={2024},
  eprint={2407.21783},
  archivePrefix={arXiv},
  primaryClass={cs.AI},
  url={https://arxiv.org/abs/2407.21783},
}

@inproceedings{wei2022chain,
  title={Chain-of-Thought Prompting Elicits Reasoning in Large Language Models},
  author={Wei, Jason and Wang, Xuezhi and Schuurmans, Dale and Bosma, Maarten and Xia, Fei and Chi, Ed and Le, Quoc and Zhou, Denny and others},
  booktitle={Advances in Neural Information Processing Systems 35 (NeurIPS 2022)},
  volume={35},
  pages={24824--24837},
  year={2022},
  address={New Orleans, USA},
  publisher={Curran Associates, Inc.},
}

@inproceedings{hsieh2023distilling,
  title={Distilling Step-by-Step! Outperforming Larger Language Models with Less Training Data and Smaller Model Sizes},
  author={Hsieh, Cheng-Yu and Li, Chun-Liang and Yeh, Chih-Kuan and Nakhost, Hootan and Fujii, Yasuhisa and Ratner, Alex and Krishna, Ranjay and Lee, Chen-Yu and Pfister, Tomas},
  booktitle={Findings of the Association for Computational Linguistics: ACL 2023},
  pages={8003--8017},
  year={2023},
  address={Toronto, Canada},
  publisher={Association for Computational Linguistics},
}

@misc{fan2025minirag,
  title={MiniRAG: Towards Extremely Simple Retrieval-Augmented Generation},
  author={Fan, Tianyu and Wang, Jingyuan and Ren, Xubin and Huang, Chao},
  year={2025},
  eprint={2501.06713},
  archivePrefix={arXiv},
  primaryClass={cs.AI},
  url={https://arxiv.org/abs/2501.06713},
}

@inproceedings{guo2024lightrag,
  title={LightRAG: Simple and Fast Retrieval-Augmented Generation},
  author={Guo, Zirui and Xia, Lianghao and Yu, Yanhua and Ao, Tu and Huang, Chao},
  booktitle={Findings of the Association for Computational Linguistics: EMNLP 2025},
  pages={10746--10761},
  year={2025},
  address={Suzhou, China},
  publisher={Association for Computational Linguistics},
  url={https://aclanthology.org/2025.findings-emnlp.568/},
}

@article{izacard2023atlas,
  title={Atlas: Few-shot learning with retrieval augmented language models},
  author={Izacard, Gautier and Lewis, Patrick and Lomeli, Maria and Hosseini, Lucas and Petroni, Fabio and Schick, Timo and Dwivedi-Yu, Jane and Joulin, Armand and Riedel, Sebastian and Grave, Edouard},
  journal={Journal of Machine Learning Research},
  volume={24},
  number={251},
  pages={1--43},
  year={2023}
}

@misc{lee2024disentangling,
  title={Disentangling Questions from Query Generation for Task-Adaptive Retrieval},
  author={Lee, Yoonsang and Kim, Minsoo and Hwang, Seung-won},
  year={2024},
  eprint={2409.16570},
  archivePrefix={arXiv},
  primaryClass={cs.CL},
  url={https://arxiv.org/abs/2409.16570},
}

@inproceedings{lewis2020retrieval,
  title={Retrieval-Augmented Generation for Knowledge-Intensive NLP Tasks},
  author={Lewis, Patrick and Perez, Ethan and Piktus, Aleksandra and Petroni, Fabio and Karpukhin, Vladimir and Goyal, Naman and K{\"u}ttler, Heinrich and Lewis, Mike and Yih, Wen-tau and Rockt{\"a}schel, Tim and Riedel, Sebastian and Kiela, Douwe},
  booktitle={Advances in Neural Information Processing Systems 33 (NeurIPS 2020)},
  volume={33},
  pages={9459--9474},
  year={2020},
  address={Vancouver, Canada},
  publisher={Curran Associates, Inc.},
}

@inproceedings{trivedi2023interleaving,
  title={Interleaving Retrieval with Chain-of-Thought Reasoning for Knowledge-Intensive Multi-Step Questions},
  author={Trivedi, Harsh and Balasubramanian, Niranjan and Khot, Tushar and Sabharwal, Ashish},
  booktitle={Proceedings of the 61st Annual Meeting of the Association for Computational Linguistics (Volume 1: Long Papers)},
  pages={10014--10037},
  year={2023},
  address={Toronto, Canada},
  publisher={Association for Computational Linguistics},
  url={https://aclanthology.org/2023.acl-long.557/},
}

@misc{wang2024rat,
  title={RAT: Retrieval Augmented Thoughts Elicit Context-Aware Reasoning in Long-Horizon Generation},
  author={Wang, Zihao and Liu, Anji and Lin, Haowei and Li, Jiaqi and Ma, Xiaojian and Liang, Yitao},
  year={2024},
  eprint={2403.05313},
  archivePrefix={arXiv},
  primaryClass={cs.CL},
  url={https://arxiv.org/abs/2403.05313},
}

@inproceedings{mavi2023retrieval,
  title={Retrieval-Augmented Chain-of-Thought in Semi-structured Domains},
  author={Mavi, Vaibhav and Saparov, Abulhair and Zhao, Chen},
  booktitle={Proceedings of the Natural Legal Language Processing Workshop 2023 (NLLP 2023)},
  pages={178--191},
  year={2023},
  address={Singapore},
  publisher={Association for Computational Linguistics},
  url={https://aclanthology.org/2023.nllp-1.18/},
}

@article{li2024rt,
  title={RT: a Retrieving and Chain-of-Thought framework for few-shot medical named entity recognition},
  author={Li, Mingchen and Zhou, Huixue and Yang, Han and Zhang, Rui},
  journal={Journal of the American Medical Informatics Association},
  volume={31},
  number={9},
  pages={1929--1938},
  year={2024},
  publisher={Oxford University Press}
}

@inproceedings{ma2023chain,
  title={Chain of Thought with Explicit Evidence Reasoning for Few-shot Relation Extraction},
  author={Ma, Xilai and Li, Jing and Zhang, Min},
  booktitle={Findings of the Association for Computational Linguistics: EMNLP 2023},
  pages={2334--2352},
  year={2023},
  address={Singapore},
  publisher={Association for Computational Linguistics},
  url={https://aclanthology.org/2023.findings-emnlp.153/},
}

@misc{chen2025towards,
  title={Towards Reasoning Era: A Survey of Long Chain-of-Thought for Reasoning Large Language Models},
  author={Chen, Qiguang and Qin, Libo and Liu, Jinhao and Peng, Dengyun and Guan, Jiannan and Wang, Peng and Hu, Mengkang and Zhou, Yuhang and Gao, Te and Che, Wanxiang},
  year={2025},
  eprint={2503.09567},
  archivePrefix={arXiv},
  primaryClass={cs.AI},
  url={https://arxiv.org/abs/2503.09567},
}

@misc{wang2022selfconsistency,
  title={Self-Consistency Improves Chain of Thought Reasoning in Language Models},
  author={Wang, Xuezhi and Wei, Jason and Schuurmans, Dale and Le, Quoc and Chi, Ed and Narang, Sharan and Chowdhery, Aakanksha and Zhou, Denny},
  year={2022},
  eprint={2203.11171},
  archivePrefix={arXiv},
  primaryClass={cs.CL},
  url={https://arxiv.org/abs/2203.11171},
}

@misc{zhou2022least,
  title={Least-to-Most Prompting Enables Complex Reasoning in Large Language Models},
  author={Zhou, Denny and Sch{\"a}rli, Nathanael and Hou, Le and Wei, Jason and Scales, Nathan and Wang, Xuezhi and Schuurmans, Dale and Cui, Claire and Bousquet, Olivier and Le, Quoc},
  year={2022},
  eprint={2205.10625},
  archivePrefix={arXiv},
  primaryClass={cs.CL},
  url={https://arxiv.org/abs/2205.10625},
}

@misc{chen2022program,
  title={Program of Thoughts Prompting: Disentangling Computation from Reasoning for Numerical Reasoning Tasks},
  author={Chen, Wenhu and Ma, Xueguang and Wang, Xinyi and Cohen, William W.},
  year={2022},
  eprint={2211.12588},
  archivePrefix={arXiv},
  primaryClass={cs.CL},
  url={https://arxiv.org/abs/2211.12588},
}

@inproceedings{yao2023tree,
  title={Tree of Thoughts: Deliberate Problem Solving with Large Language Models},
  author={Yao, Shunyu and Yu, Dian and Zhao, Jeffrey and Shafran, Izhak and Griffiths, Thomas L. and Cao, Yuan and Narasimhan, Karthik},
  booktitle={Advances in Neural Information Processing Systems 36 (NeurIPS 2023)},
  volume={36},
  pages={11809--11822},
  year={2023},
  address={New Orleans, USA},
  publisher={Curran Associates, Inc.},
}

@misc{bi2024forest,
  title={Forest-of-Thought: Scaling Test-Time Compute for Enhancing LLM Reasoning},
  author={Bi, Zhenni and Han, Kai and Liu, Chuanjian and Tang, Yehui and Wang, Yunhe},
  year={2024},
  eprint={2412.09078},
  archivePrefix={arXiv},
  primaryClass={cs.CL},
  url={https://arxiv.org/abs/2412.09078},
}

@misc{pandey2025adaptive,
  title={Adaptive Graph of Thoughts: Test-Time Adaptive Reasoning Unifying Chain, Tree, and Graph Structures},
  author={Pandey, Tushar and Ghukasyan, Ara and Goktas, Oktay and Radha, Santosh Kumar},
  year={2025},
  eprint={2502.05078},
  archivePrefix={arXiv},
  primaryClass={cs.AI},
  url={https://arxiv.org/abs/2502.05078},
}

@book{newell1972human,
  title={Human Problem Solving},
  author={Newell, Allen and Simon, Herbert A.},
  year={1972},
  publisher={Prentice-Hall},
  address={Englewood Cliffs, NJ}
}

@article{cushen2012cues,
  title = {Cues to Solution, Restructuring Patterns, and Reports of Insight in Creative Problem Solving},
  journal = {Consciousness and Cognition},
  volume = {21},
  number = {3},
  pages = {1166-1175},
  year = {2012},
  author = {Cushen, Patrick J. and Wiley, Jennifer},
}

@article{ho2022people,
  title={People construct simplified mental representations to plan},
  author={Ho, Mark K. and Abel, David and Correa, Carlos G. and Littman, Michael L. and Cohen, Jonathan D. and Griffiths, Thomas L.},
  journal={Nature},
  volume={606},
  number={7912},
  pages={129--136},
  year={2022},
  publisher={Nature Publishing Group UK London}
}

@inproceedings{eckstein2021mind,
  title={How the Mind Creates Structure: Hierarchical Learning of Action Sequences},
  author={Eckstein, Maria K. and Collins, Anne G. E.},
  booktitle={Proceedings of the Annual Meeting of the Cognitive Science Society (CogSci 2021)},
  volume={43},
  pages={618--624},
  year={2021},
  address={Vienna, Austria},
  publisher={Cognitive Science Society},
}

@incollection{barto2013behavioral,
  title={Behavioral Hierarchy: Exploration and Representation},
  author={Barto, Andrew G. and Konidaris, George and Vigorito, Christopher},
  booktitle={Computational and Robotic Models of the Hierarchical Organization of Behavior},
  pages={13--46},
  year={2013},
  publisher={Springer},
  address={Berlin, Heidelberg},
}

@inproceedings{zhang2023reasoning,
  title={Reasoning over Hierarchical Question Decomposition Tree for Explainable Question Answering},
  author={Zhang, Jiajie and Cao, Shulin and Zhang, Tingjian and Lv, Xin and Li, Juanzi and Hou, Lei and Shi, Jiaxin and Tian, Qi},
  booktitle={Proceedings of the 61st Annual Meeting of the Association for Computational Linguistics (Volume 1: Long Papers)},
  pages={14556--14570},
  year={2023},
  address={Toronto, Canada},
  publisher={Association for Computational Linguistics},
  url={https://aclanthology.org/2023.acl-long.814/},
}

@misc{openai2024gpt4o,
  author = {{OpenAI}},
  title = {{GPT-4o} System Card},
  howpublished = {\url{https://openai.com/index/gpt-4o-system-card/}},
  year = {2024}
}

@inproceedings{zheng2025barexam,
  author    = {Zheng, Lucia and Guha, Neel and Arifov, Javokhir and Zhang, Sarah and Skreta, Michal and Manning, Christopher D. and Henderson, Peter and Ho, Daniel E.},
  title     = {A Reasoning-Focused Legal Retrieval Benchmark},
  booktitle = {Proceedings of the 2025 Symposium on Computer Science and Law (CS\&Law '25)},
  year      = {2025},
  address   = {Munich, Germany},
  publisher = {ACM},
  doi       = {10.1145/3709025.3712219},
}

@misc{wang2025rare,
  author    = {Wang, Zhengren and Yu, Jiayang and Ma, Dongsheng and Chen, Zhe and Wang, Yu and Li, Zhiyu and Xiong, Feiyu and Wang, Yanfeng and E, Weinan and Tang, Linpeng and Zhang, Wentao},
  title     = {{RARE}: Retrieval-Augmented Reasoning Modeling},
  year      = {2025},
  eprint    = {2503.23513},
  archivePrefix = {arXiv},
  primaryClass  = {cs.CL},
  url       = {https://arxiv.org/abs/2503.23513},
}

@inproceedings{zhang2025imprag,
  author    = {Zhang, Wenzheng and Lin, Xi Victoria and Stratos, Karl and Yih, Wen-tau and Chen, Mingda},
  title     = {{ImpRAG}: Retrieval-Augmented Generation with Implicit Queries},
  booktitle = {Findings of the Association for Computational Linguistics: EMNLP 2025},
  year      = {2025},
  address   = {Suzhou, China},
  publisher = {Association for Computational Linguistics},
}


\end{document}